\documentclass[conference]{IEEEtran}
\IEEEoverridecommandlockouts
\usepackage{cite}
\usepackage{amsmath,amssymb,amsfonts}
\usepackage{algorithmic}
\usepackage{graphicx}
\usepackage{textcomp}
\usepackage{xcolor}
\usepackage{caption}
\usepackage{subcaption}
\usepackage{times}
\usepackage{mwe}
\usepackage{multirow}

\usepackage{xcolor}
\usepackage[switch]{lineno}
\usepackage{url}
\usepackage{lipsum}
\usepackage{graphicx}
\usepackage{float}
\usepackage{amsfonts}
\usepackage{amsmath}
\usepackage[ruled,noline,linesnumbered,noend]{algorithm2e}
\usepackage{booktabs}
\usepackage[bookmarks=false]{hyperref}
\usepackage{cite}
\usepackage{array}
\usepackage{graphbox}

\usepackage{float}

\usepackage{tikz}

\def\BibTeX{{\rm B\kern-.05em{\sc i\kern-.025em b}\kern-.08em
    T\kern-.1667em\lower.7ex\hbox{E}\kern-.125emX}}
    
\begin{document}

\title{\LARGE Sparse-View CT Reconstruction using Recurrent Stacked Back Projection\\
\thanks{Wenrui Li, and Charles A. Bouman are with the School of Electrical and Computer Engineering, Purdue University, West Lafayette, IN, USA (e-mail: \href{mailto:li3120@purdue.edu}{li3120@purdue.edu} , \href{mailto:bouman@purdue.edu}{bouman@purdue.edu}),
Gregery T. Buzzard is with the Department of Mathematics, Purdue University, West Lafayette, IN, USA (email: \href{mailto:buzzard@purdue.edu}{buzzard@purdue.edu}).
Li and Bouman were partially supported by the US Dept. of Homeland Security, S\&T Directorate, under Grant Award 2013-ST-061-ED0001.
Buzzard was partially supported by the NSF under grant award  CCF-1763896.
The views and conclusions contained in this document are those of the authors and should not be interpreted as necessarily representing the official policies, either expressed or implied, of the U.S. Department of Homeland Security.
}
}
\makeatletter
\author{\IEEEauthorblockN{Wenrui Li}
\IEEEauthorblockA{\textit{Electrical and Computer Engineering} \\
\textit{Purdue University}\\
West Lafayette, IN, United States \\
li3120@purdue.edu}
\and
\IEEEauthorblockN{Gregery T. Buzzard}
\IEEEauthorblockA{\textit{Mathematics} \\
\textit{Purdue University}\\
West Lafayette, IN, United States \\
buzzard@purdue.edu}
\and
\IEEEauthorblockN{Charles A. Bouman}
\IEEEauthorblockA{\textit{Electrical and Computer Engineering} \\
\textit{Purdue University}\\
West Lafayette, IN, United States \\
bouman@purdue.edu}

}

\maketitle

\begin{abstract}
Sparse-view CT reconstruction is important in a wide range of applications due to limitations on cost, acquisition time, or dosage. However, traditional direct reconstruction methods such as filtered back-projection (FBP) lead to low-quality reconstructions in the sub-Nyquist regime. In contrast, deep neural networks (DNNs) can produce high-quality reconstructions from sparse and noisy data, e.g. through post-processing of FBP reconstructions, as can model-based iterative reconstruction (MBIR), albeit at a higher computational cost. 

In this paper, we introduce a direct-reconstruction DNN method called Recurrent Stacked Back Projection (RSBP) that uses sequentially-acquired backprojections of individual views as input to a recurrent convolutional LSTM network. The SBP structure maintains all information in the sinogram, while the recurrent processing exploits the correlations between adjacent views and produces an updated reconstruction after each new view. We train our network on simulated data and test on both simulated and real data and demonstrate that RSBP outperforms both DNN post-processing of FBP images and basic MBIR, with a lower computational cost than MBIR.
\end{abstract}

\begin{IEEEkeywords}
Sparse-view CT, CT reconstruction, deep learning, Recurrent Stacked Back Projection, LSTM
\end{IEEEkeywords}

\section{Introduction} \label{Introduction}
In many applications of computed tomography (CT), the number of views that can be acquired is limited due to cost, acquisition time, hardware constraints, or allowable X-ray dosage.  This motivates the challenging problem of reconstructing high quality images from sparse-view CT data.
Traditional direct reconstruction algorithms such as filtered back projection (FBP) depend on the assumption of Nyquist sampling, hence can produce severe artifacts when given sparse-view data \cite{chen2008prior}.
Iterative reconstruction algorithms such as model-based iterative reconstruction (MBIR) can dramatically improve the quality of sparse view reconstructions, but at the cost of higher computation \cite{CTMeeting2012Kisner, thibault2007three, Venkat2013plug}.

More recently, the use of deep neural networks (DNNs) has emerged as a fundamentally new approach to tomographic reconstruction \cite{Jin2017depp} with the advantages that a) it can greatly reduce computation, and b) given sufficient training data, it can be trained to incorporate complex prior information\cite{ravishankar_image_2019}.
One way to use DNNs for CT reconstruction is to preprocess sinogram data \cite{lee2017view,lee_deep-neural-network_2019,ghani2018deep} and then apply a standard reconstruction; however, this approach does not fully utilize the capabilities of DNNs in the image domain.  A second approach uses DNNs for post-processing of an image reconstructed using a simple method such as FBP\cite{Asilomar2018Ziabari,han2018framing,zhang2018sparse}; this approach leverages the capabilities of DNNs in the image domain but does not have full access to all sinogram data.  Finally, direct reconstruction from the sinogram domain using DNNs as in \cite{ravishankar_image_2019} is very memory and computationally expensive \cite{zhu_image_2018}.

In this paper, we introduce a direct-reconstruction DNN method called Recurrent Stacked Back Projection (RSBP) that uses single-view backprojections as sequential input to a recurrent convolutional LSTM network, whose output is then post-processed by a U-Net.
The Stacked Back Projection (SBP) maintains all information in the sinogram \cite{ye2018deep}, while the recurrent processing exploits the correlations between adjacent views and produces an updated reconstruction after each new view.  
We train our network on simulated data and test on both simulated and real data and demonstrate that RSBP outperforms similar DNN post-processing of FBP images or SBP tensors and also outperforms basic MBIR, with a lower computational cost than MBIR.

\begin{figure}[!t]
\centerline{\includegraphics[width=0.95\linewidth]{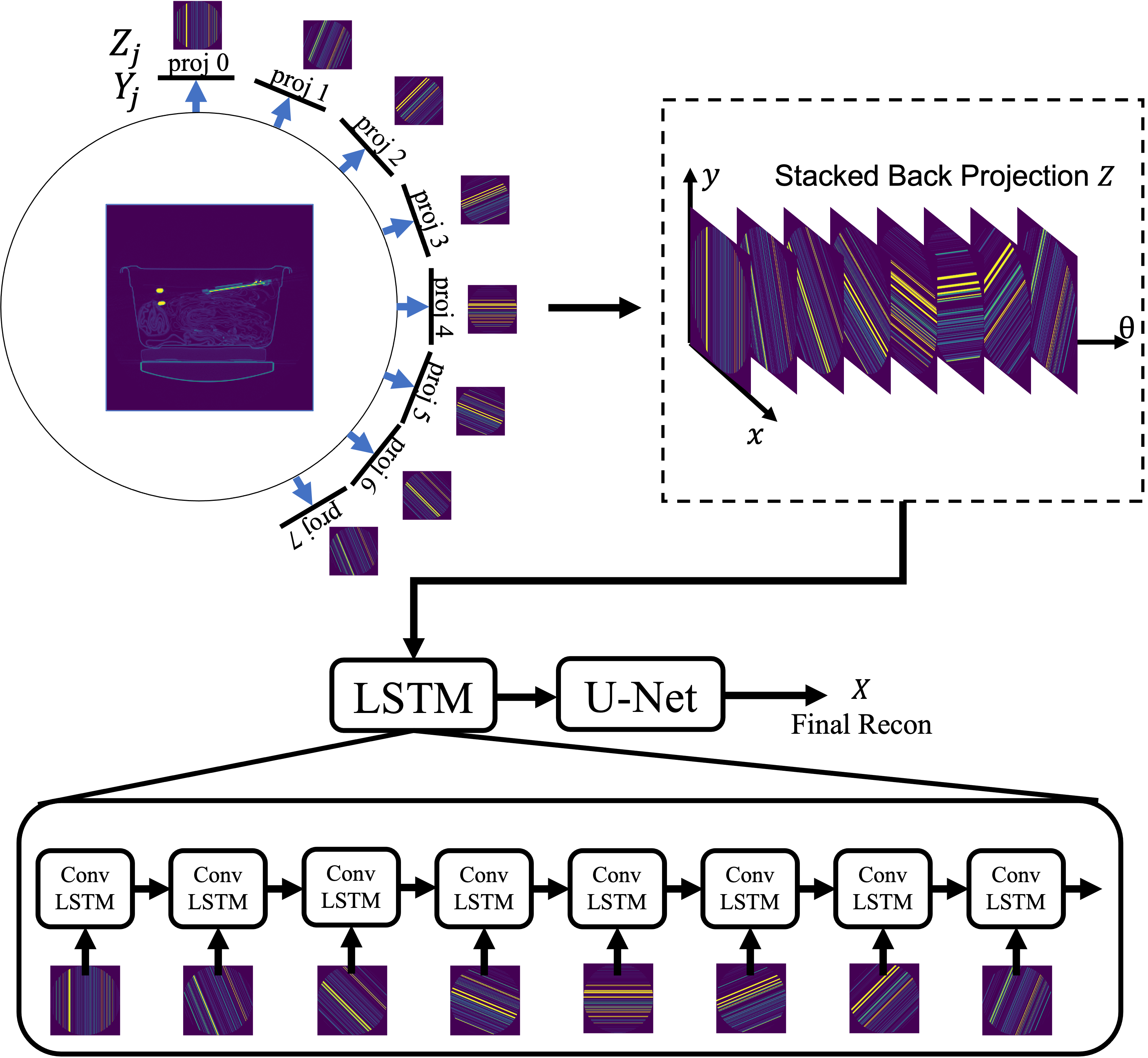}}
\caption{\textbf{Conceptual view of RSBP architecture:} 
a) The single-view projections $Z_j$ are back-projected from a single projection $y_j$, then collected into a tensor called the Stacked Back Projection. b) Each single-view back-projection is input to the LSTM sequentially, then further processed by a U-Net.}
\label{fig:sbp}
\end{figure}

\section{Recurrent Stacked Back Projection} \label{Recur_BP}

In Fig.~1, we illustrate the network architecture of our RSBP method, in which individual backprojections are stacked in the image domain and fed sequentially to a convolutional LSTM, whose image output is then fed to a U-Net for final reconstruction. 
In section \ref{ssec:Recur_BP_tensor}, we introduce the SBP tensor, which contains all information from the sparse-view sinogram.
In section \ref{ssec:convlstm}, we introduce each component of our proposed network in more detail.

\subsection{Stacked Back Projection Tensor}\label{ssec:Recur_BP_tensor}
We consider a CT reconstruction problem in which an unknown image $x\in \mathbb{R}^{N \times N}$ is observed via $N$ detectors and $M$ views to yield noisy measurements $y \in \mathbb{R}^{N\times M}$. 
Each column vector $y_j \in \mathbb{R}^{N}$ approximates the integral of attenuation along the line with angle $\theta_j$ according to the model
\begin{equation}
    y_j = A_{j}x+w_j,
\end{equation}
where $A_{j}$ is the Radon transform operator with angle $\theta_j = \frac{j\pi}{M}$ for $j=\{0, 1, .., M-1\}$, and $w_j\in \mathbb{R}^{N}$ represents signal-dependent noise by $w_j \sim N(0, \text{diag}(\frac{1}{\lambda_0} \exp(A_j x)))$, where $\lambda_0$ is the empty scan photon count.  

To obtain the SBP tensor, we first apply single-view FBP to each measured projection to obtain
\begin{equation}
    Z_j=A_{j}^{\dagger}(y_j),
\end{equation}
where $A_{j}^{\dagger}$ is the inverse Radon transform using a
single-view FBP operator with angle $\theta_j$.
As in \cite{ye2018deep}, the images $Z_j \in\mathbb{R}^{N \times N}$ are then stacked to form a rank 3 SBP tensor, in our case with dimensions $N\times N \times M$:
\begin{equation}
    \boldsymbol{Z}=\{Z_1;Z_2;...;Z_M\}.
\end{equation}

In contrast to an FBP image, the SBP tensor contains all information from sparse-view CT measurements. But like the FBP, the SBP converts the complex spatial correlations in the sinogram back to the image domain, which allows for the use of efficient processing by convolutional neural networks (CNNs), as shown in Fig.~\ref{fig:diff_models}(b).

\subsection{Convolutional LSTM Network with SBP}\label{ssec:convlstm}
\begin{figure}[!t]
\centerline{\includegraphics[width=0.9\linewidth]{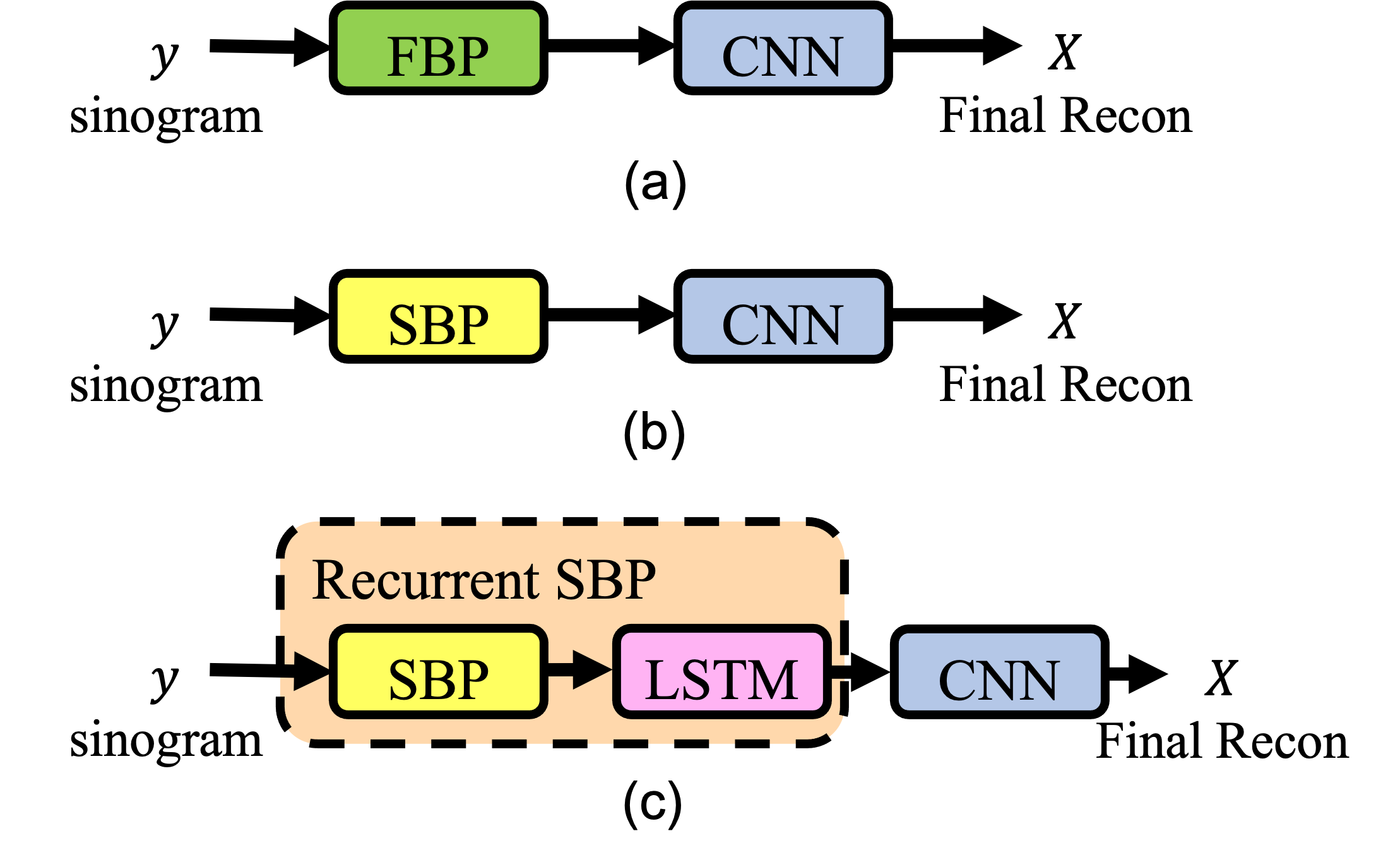}}
\caption{\textbf{Evolution towards RSBP:}
a) CNN post-processing of FBP image; b) simultaneous CNN post-processing of SBP tensor; c) sequential input of SBP backprojections to convolutional LSTM followed by CNN postprocessing.}
\label{fig:diff_models}
\end{figure}

Two problems with the SBP as used in \cite{ye2018deep} are that it requires advance knowledge of the number views and requires that all the single-view backprojections be stored in memory.  
That is, post-processing the SBP with a CNN as in Fig.~\ref{fig:diff_models}(b) treats each single-view back projection as an independent channel, like RGB channels in a color image, and processes these channels simultaneously.

To address these problems, we introduce a convolutional LSTM\cite{shi2015convolutional} to process these single-view backprojections sequentially.  Most applications of recurrent networks such as the LSTM involve time-dependent measurements, such as speech processing, while most CT reconstruction methods use all available views simultaneously.  However, as shown in Fig.~\ref{fig:sbp}, the SBP not only contains all single-view backprojections but also is sequentially acquired. 
This sequential acquisition process and the two problems identified above lead naturally to propose a recurrent network.  
The advantage of a convolutional LSTM in this context is that it has the spatial invariance of a CNN and can be trained with sequentially acquired data. We call this approach recurrent stacked back-projection (RSBP), as shown in Fig.~\ref{fig:diff_models}(c).

\begin{figure}[!t]
\centerline{\includegraphics[width=0.95\linewidth]{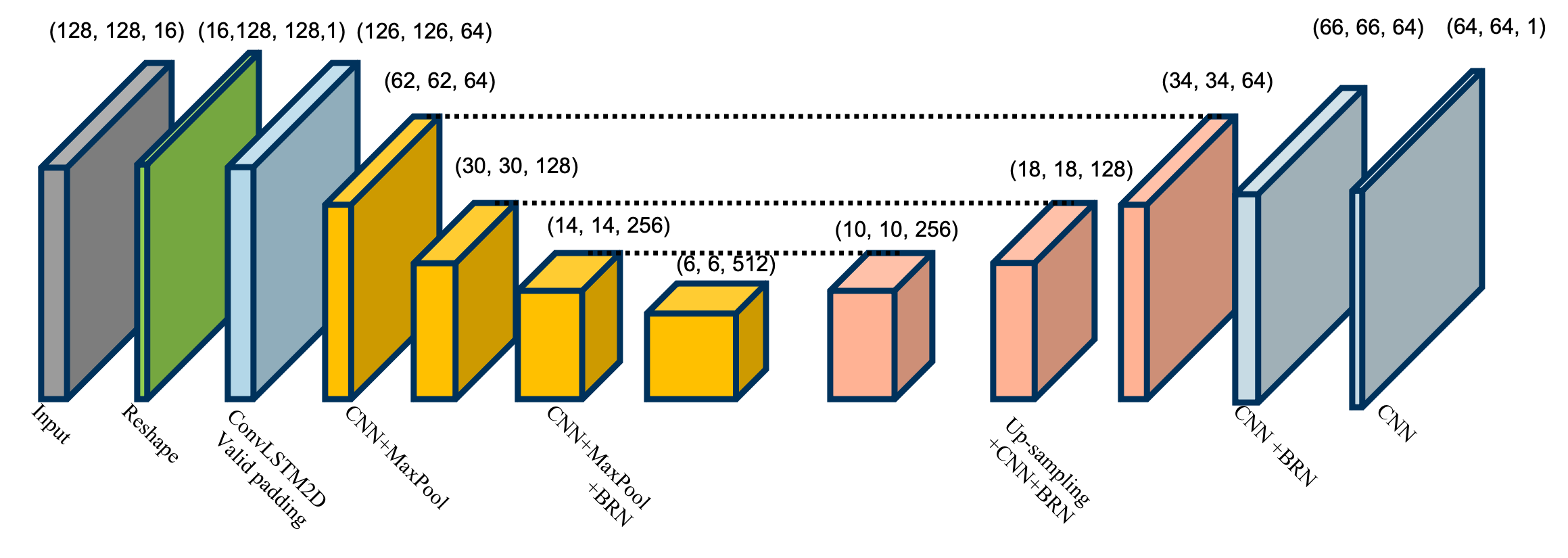}}
\caption{\textbf{RSBP architecture:}
The input SBP is fed sequentially to a convolutional LSTM, with output then processed by a U-Net using valid padding.  With valid padding, the output image is smaller than the input image, so we use only the central region of the corresponding ground truth image to calculate loss functions and quality metrics.}
\label{fig:rsbp_model}
\end{figure}

\subsection{Network architectures}\label{ssec:network} 
The full RSBP-CNN architecture is illustrated in Fig.~\ref{fig:rsbp_model}, which shows the initial processing by the LSTM followed by a U-Net\cite{ronneberger2015u}.
We first reshape the SBP tensor from (Rows, Columns, Views) to (Views, Rows, Columns, Channels) and then feed this to the convolutional LSTM layer, with kernel size $3 \times 3$ for both the input transformations and recurrent transformations.
From the LSTM we obtain a (Rows, Columns, Channels) tensor and pass it to a U-Net.
We use batch renormalization in most of the downsampling and upsampling blocks for better convergence and stability of the network\cite{ioffe2015batch, ioffe2017batch}.
In the final layer, we use convolution with a $3 \times 3$ kernel to generate a single output image.

\section{Implementation}
We implement and train RSBP-CNN using Keras with a loss function described below.  We use previously reconstructed full-view CT data to serve as ground truth;  we use 8 or 16 simulated projection views to serve as input to RSBP.  We evaluate the method on both simulated sinograms and real sinograms.

\subsection{Dataset Generation}  \label{sec:dataset}
We generate training and testing data using the ALERT Task Order 4 dataset\cite{alertto4}, which consists of 188 full-view sinograms and high-quality 3D reconstructions of packed suitcases using data from an Imatron Scanner.  
First, we split the data into 153 3D reconstructions for training and 35 for testing.  From each reconstruction, we randomly select 28 slices to yield 4284 slices for training and 980 slices for testing.
\begin{table}[!t]
\centering
\caption{Setup for simulated X-ray data experiment}
\begin{tabular}{rl}
\hline
Projection Geometry: & Parallel beam geometry \\ 
Number of Views: & 8 or 16 equi-spaced angles in $[0,\pi]$ \\
Pixel Pitch $p$: & 0.186 cm \\
FOV: & 47.616 cm \\
Water X-ray density $\mu$: & 0.17 $cm^{-1}$ ($\sim$100 keV) \\
Photon dosage per proj ($\lambda_0$): & 1600 \\ \hline
\end{tabular}

\label{Table:1}
\end{table}
Next, we generate simulated projections using the scaling and noise modeling approach described in \cite{buzzard_bouman_2019}. For each view angle $\theta_j$ we use the forward model
\begin{equation} \label{Equ:forward} 
y_j=\left(\mu p\right) A_j \left(\frac{x}{1000}\right)+w_j \; ;
\end{equation}
here division by 1000 scales the image from (modified) Hounsfield units\footnote{We use modified Hounsfield units in which air is 0 and water is 1000.} to units in which water is 1 and air is 0, $A_j$ is the raw radon transform for $\theta_j$, $\mu$ and $p$ are the water X-ray density and pixel pitch, respectively, and $w_j \sim N(0, R)$ with $R$ a diagonal matrix with entries $\frac{\exp \left\{A_j x\right\}}{\lambda_{0}}$ to model measurement noise.  The physical constants are defined in Table~\ref{Table:1}. 

From the scaled simulated sinogram $y$, we  apply single-view back projection and inverse scaling, then stack as in  Fig.~\ref{fig:sbp} to generate the SBP. 
This process results in pairs of ground truth images and corresponding simulated SBP tensors in Hounsfield units.  The SBP tensor is scaled back to units in which water is 1 before sending to the NN, then scaled back to Hounsfield units before evaluating the loss function.

\subsection{Loss Function}

For reconstructing an X-ray CT image for security applications, the range from $0$ to $2000$ Hounsfield units (HU) (air=0 HU, water=1000 HU) is critical. 
Therefore, we use a modified MSE loss to train our models:
\begin{equation}
L_{M S E}=\|f(x_{true})-f(\hat{x})\|_2^{2},
\end{equation}
where $x_{true}$ is the ground truth (cropped to the valid region as described in Fig.~\ref{fig:rsbp_model}), $\hat{x}$ is the reconstructed image, $f(x)=\frac{x}{|x|+2000}$, and both $x_{true}$ and $\hat{x}$ are in Hounsfield units.

\subsection{Training Process}

We implemented all NN models in Tensorflow Keras and used one NVIDIA Tesla V100 GPU for training. 
In order to reduce memory requirements and increase batch size during training \cite{lee2018deep}, we used randomly cropped patches of size $128 \times 128$ from the SBP tensors (original image size $320 \times 320$) as input and calculated the loss between the output ($64 \times 64$) and the center of the corresponding patches from the ground truth images.
In inference mode, we applied the trained model on full image slices.
We used the Adam optimizer with learning rate = $0.0002$. For each model, we trained for 120 epochs 
and shuffled the training dataset after each epoch. 


\begin{figure*}[!t]
\centerline{\includegraphics[width=0.95\linewidth]{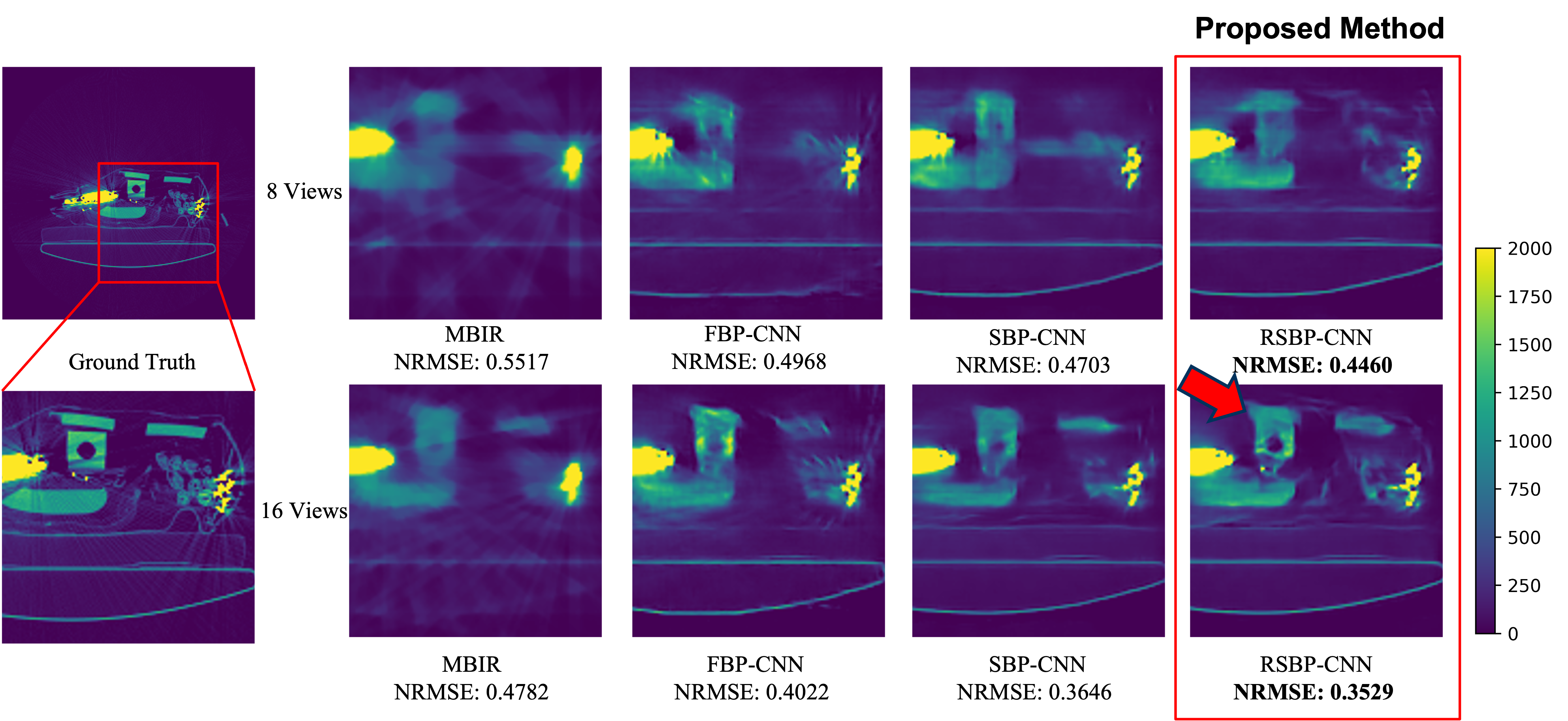}}
\caption{\textbf{Comparisons of multiple reconstruction algorithms with simulated sinogram:} \label{fig:SimulationResults}
  {\em Left 2 panels:} Ground truth image used to generate simulated sinogram data and a cropped detail used for display comparisons.  {\em Right panels: }  Reconstructions using MBIR, FBP-CNN, SBP-CNN and RSBP-CNN.  The top row uses 8 views and the bottom row uses 16 views.  The NRMSE is for the full image rather than the displayed cropped image. 
The display range is from $0$ (air) to $2000$ Hounsfield units (HU). 
The RSBP-CNN reconstruction has more detail and lower NRMSE relative to the other methods.}
\end{figure*}

\section{Experiments}
We compare our proposed RSBP-CNN model with several alternatives, including (a) MBIR\cite{wang2016high} using the QGGMRF prior, (b) FBP-CNN, and (c) SBP-CNN. 
FBP-CNN and SBP-CNN, illustrated in Fig.~\ref{fig:diff_models}, both use the architecture in Fig.~\ref{fig:rsbp_model} with a convolutional layer in place of the LSTM; we train them with the same strategy use for RSBP.

We examine reconstructions using either 8 views or 16 views.  In each of these cases we trained each of FBP-CNN, SBP-CNN, and RSBP-CNN for a total of 6 trained networks. We tested these DNN methods and MBIR on  simulated data as described above and on real sinogram data of 9 different 3D objects from the ALERT Task Order 3 dataset\cite{alertto3}.

Since we apply the raw inverse radon transform on real sinogram data, we apply inverse scaling to the SBP as before.  The parameters are the same as in Table~\ref{Table:1} except that now $\mu$ is 0.159 $cm^{-1}$ ($\sim$130 keV).

In each reconstruction, we computed NRMSE (normalized root mean square error) between reconstruction and ground truth using
\begin{equation}
 NRMSE = \frac{\| x_{true}-\hat{x}\|_{2} }{\| x_{true}\|_{2} } \;.
\end{equation}
For results using real sinogram data, we use an MBIR reconstruction with 720 views as ground truth.  

Fig.~\ref{fig:SimulationResults} shows reconstructions with each of the 4 methods using each of 8-view and 16-view sinogram data.  The images and NRMSE show that RSBP-CNN improves over SBP-CNN, which improves over FBP-CNN, in both cases for both 8-view and 16-view sinogram data. RSBP-CNN eliminates most of the streak artifacts seen in the MBIR and FBP-CNN reconstructions and preserves more details than SBP-CNN. 

The improvement from using SBP is consistent with results of \cite{ye2018deep} and the fact that the SBP includes all sinogram data, which is not true of the FBP.  
The improvement from SBP-CNN to RSBP-CNN is more subtle; in principle SBP-CNN has all the information need to perform as well as RSBP-CNN.  We hypothesize that sequential processing of individual backprojections promotes better use of correlations between adjacent views; in addition, there may also be interactions between our training scheme and the reduced memory requirements of RSBP relative to simultaneous processing of the SBP.

\begin{table*}[!t]
\centering
\caption{Mean / standard deviation of NRMSE for different reconstruction algorithms on multiple datasets}
\label{Table:3}
\begin{tabular}{|l|l|c|c|c|c|}
\hline
\multicolumn{2}{|c|}{\textbf{Method}} &
  \textbf{MBIR} &
  \textbf{FBP-CNN} &
  \textbf{SBP-CNN} &
  \textbf{RSBP-CNN} \\ \hline
\multirow{2}{*}{\begin{tabular}[c]{@{}l@{}}Simulated\\ sinogram\end{tabular}} &
  8-View &
  0.567 / 0.084 &
  0.478 / 0.084 &
  0.454 / 0.081 &
  \textbf{0.428 / 0.084} \\ \cline{2-6} 
 & 16-View & 
  0.498 / 0.080& 
  0.381 / 0.076 & 
  0.375 / 0.070 & 
  \textbf{0.335 / 0.070} \\ \hline
\multirow{2}{*}{\begin{tabular}[c]{@{}l@{}}Real\\ sinogram\end{tabular}} &
  8-View &
  0.553 / 0.088&
  0.590 / 0.112&
  0.519 / 0.085&
  \textbf{0.506 / 0.092} \\ \cline{2-6} 
 & 16-View & 
  0.462 / 0.087& 
  0.411 / 0.089 & 
  0.405 / 0.077& 
  \textbf{0.368 / 0.085} \\ \hline
\end{tabular}
\end{table*}

\begin{figure*}[!t]

\centerline{\includegraphics[width=0.95\linewidth]{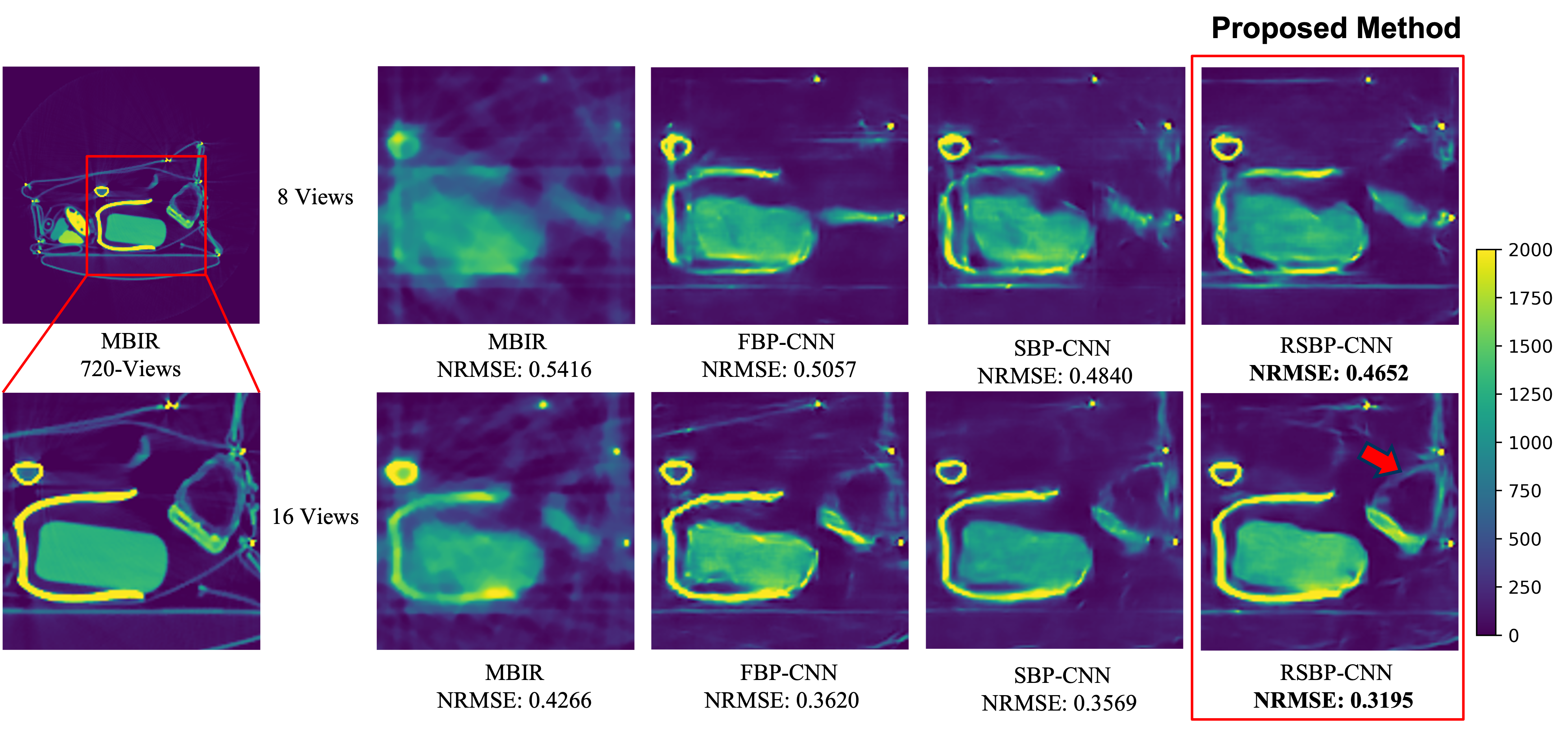}}
\caption{\textbf{Comparisons of multiple reconstruction algorithms with real sinogram:} \label{fig:RealResults}
{\em Left 2 panels:} Ground truth image generated from full view real sinogram data using MBIR and a cropped detail used for display comparisons.  {\em Right panels: }  Reconstructions using MBIR, FBP-CNN, SBP-CNN, and RSBP-CNN.  The top row uses 8 views and the bottom row uses 16 views.  The NRMSE is for the full image rather than the displayed cropped image. 
The display range is from $0$ (air) to $2000$ Hounsfield units (HU). 
The RSBP-CNN reconstruction has clearer edges and lower NRMSE relative to the other methods.} 
\end{figure*}

Fig.~\ref{fig:RealResults} repeats the experiment of Fig.~\ref{fig:SimulationResults} using real sinogram data.  In this case we use a full-view MBIR reconstruction as ground truth to compute the NRMSE.  Once again we see that RSBP improves over SBP, which improves over FBP.

In Table~\ref{Table:3}, we see that the performance gains from using RSBP and SBP are consistent across multiple trials.  That table shows the mean and standard deviation of NRMSE over the simulated test data (980 scans) and the real data (900 scans).  In all cases, there is a signifcant drop in NRMSE when going from FBP-CNN to SBP-CNN and again from SBP-CNN to RSBP-CNN.  Hence RSBP shows real promise for truly sparse-view reconstruction.

\section{Conclusion and Future work}
In this paper, we introduced a novel, direct, deep learning algorithm RSBP-CNN that combines SBP, LSTM, and U-Net to reconstruct CT images from 8 or 16 view sinogram data. 
Our method couples the representational efficiency of the SBP as input with the computational and inferential efficiency of a convolutional LSTM. 
The use of an LSTM brings the ability to learn embedded order information from the sequentially-acquired SBP, thus enabling better reconstruction than the same network with a CNN in the first layer;  LSTM also yields a reduction in memory requirements due to the sequential processing of back projections in the SBP. 
Our proposed RSBP-CNN outperforms other methods in terms of visual and metric comparisons on both simulated and real sinogram data. Future work will investigate modifications to promote the ability of a single RSBP-CNN to produce reconstructions without specifying in advance the number of views.
\bibliographystyle{IEEEtran}
\bibliography{references}

\end{document}